\documentstyle[preprint,aps]{revtex}

\begin{document}


\title{Stability and collapse of   a coupled  Bose-Einstein
condensate}

\author{Sadhan K. Adhikari}
\address{Instituto de F\'{\i}sica Te\'orica, Universidade Estadual
Paulista, 01.405-900 S\~ao Paulo, S\~ao Paulo, Brazil\\}

\date{\today}
\maketitle
\begin{abstract}

The dynamics of a coupled Bose-Einstein condensate involving
trapped  atoms in two quantum states is studied using the time-dependent
Gross-Pitaevskii
equation including an interaction which can transform atoms from one state
to the other. We find interesting oscillation of the number of atoms in
each of the states. For all repulsive
interactions, stable condensates are formed. 
When some of the atomic interactions are attractive, the possibility of
collapse is
studied 
by including
an absorptive contact interaction and a quartic three-body recombination
term. One or both components of the condensate may undergo collapse when
one or more of the nonlinear terms are attractive in nature.

{\bf PACS Number(s): 03.75.Fi}

\end{abstract}


\newpage
\section{Introduction}

Recent experimental success \cite{a1,b1,c1,d1,e1} in Bose-Einstein
condensation
(BEC) at ultralow temperature in dilute trapped alkali metal and hydrogen
atoms has intensified theoretical studies of the inhomogeneous,
weakly interacting dilute Bose gas and its condensation
\cite{f1,g1,h1,i1,j1,k1,l1,m1,n1,o1,p1,q1}.
At zero temperature such a system is supposed to be fully condensed and
described by the mean-field Gross-Pitaevskii (GP)  equation
\cite{l1,r1,s1}.

Two most interesting phenomena in BEC are the collapse for attractive
interaction and the formation of coupled atomic condensate. For attractive
atomic interaction \cite{d1,e1,f1,g1}, the condensate is stable for a
maximum
critical number of atoms. In the presence of an external source of atoms
when the number increases beyond the critical number, due to interatomic
attraction the radius of BEC tends to zero and the central density tends
to infinity. Consequently, the condensate collapses emitting atoms until
the number of atoms is reduced below the critical number and a stable
configuration is reached.  Then the condensate can grow again and  a
series of collapses can take place, which was observed experimentally in
the BEC of $^7$Li with attractive interaction \cite{d1,e1,f1,g1}.

There has also  been experimental realization of BEC involving
atoms in two different quantum states \cite{t1,u1,v1}. In one
experiment $^{87}$Rb atoms formed in the $|F=1$, $m=-1\rangle$ and $|F=2$,
$m=1\rangle$
states by the use of a laser served as two different species, where $F$
and $m$ are the total angular momentum and its projection \cite{t1,u1}.
In another experiment a coupled BEC was formed with the $^{87}$Rb atoms in
the $|F=1, m=-1\rangle$ and $|F=2, m=2\rangle$ states \cite{l1,v1}.
It
is possible to
use the same magnetic trap to confine atoms in two quantum  states and
this makes these experiments technically simpler compared
to a realization of BEC with two different types of atoms requiring two
independent trapping mechanisms.

Here we study theoretically the dynamics of  a coupled BEC composed of two
quantum states 1 and 2 of an atom using the coupled time-dependent GP
equation. We include an interaction term in the Hamiltonian that allows
for a coherent boson-number-conserving nondissipative transition of atoms
from one quantum state to another. This interaction simulates the action
of a laser which can been used experimentally to transform one species of
atoms to another.  This  model exhibits  rich
and novel  phenomenology  which we study using the  
numerical solution of the coupled GP equation, which
can generate
 oscillation in
the number of atoms in a specific state in the presence of the
interaction that transforms one spices of atoms to another. Also,
depending on the nature of the  atomic interactions one can have three
 possibilities of BEC: (i) stable condensate of both
components for all repulsive interactions, (ii) collapse of both 
components for
all attractive interactions, and (iii) a stable
condensate for one component and collapse in the other.

The collapse for attractive
interaction(s) in a coupled condensate 
is studied by introducing absorptive contact interactions
responsible for a growth in the atomic numbers from external source in
addition to imaginary three-body quartic interactions leading to
recombination loss \cite{f1,g1}.  In the presence of these imaginary 
interactions the GP equation does not conserve the overall
number of atoms. If the strengths
(and signs) of these imaginary interactions are properly chosen, one can
have collapse in one or both components of the condensate. 
The
solution of the GP equation could produce a growth of the condensate with
time for atom numbers less than the critical
numbers for collapse. 
Once they increase past the critical numbers, the three-body
interaction takes control and the numbers suddenly drop below the critical
 level by recombination loss signaling a collapse \cite{f1,g1}. Then
the absorptive term takes over and the individual numbers start to
increase. This continues indefinitely showing an infinite sequence of
collapses for one or both components.

We motivate this study by considering two possible atomic states 1 and 2
of $^7$Li (attractive interaction) and of $^{87}$Rb (repulsive
interaction) whenever possible. 
In the case of $^7$Li the interaction in state 1
is
taken to be attractive which is responsible for  collapse. 
We study the collapse with  different possibilities of
attraction and
repulsion between atoms in states 1 and 2 of $^7$Li.

\section{Coupled Gross-Pitaevskii Equation}

We consider the following Hamiltonian to describe the coupled BEC
allowing for the possibility of transforming atoms from one state to
another \cite{l1}
\begin{equation}
\hat H= \hat H_0+\hat V_1+\hat V_2,
\end{equation}
where the kinetic and potential energies are
\begin{equation}
\hat H_0= \int d\vec r \sum_{j=1}^2\hat  \psi_j^\dagger (\vec r) \left[
-\frac{\hbar^2}{2m}\nabla_j ^2+V^{(j)}_{\mbox{tr}}(\vec r)    
 \right]\hat \psi_j(\vec r),
\end{equation}
\begin{equation}
\hat V_1= \sum_{l,j}\frac{g_{lj}}{2}\int d\vec r 
\hat \psi_l^\dagger (\vec
r)  \hat \psi_j^\dagger ({\vec r}{}{}) 
\hat \psi_l (\vec
r)  \hat \psi_j ({\vec r}{}{}),
\end{equation}
\begin{equation}
\hat V_2= {\chi}\int d\vec r 
[\hat  \psi_1^\dagger (\vec r) \hat \psi_2 (\vec r)  +\hat  \psi_2^\dagger
(\vec r)
\hat \psi_1 (\vec r)] ,
\end{equation}
where $\hat \psi_j(\vec r)$ and  $\hat \psi^\dagger _j(\vec r)$ are  the 
field operators for annihilation and creation of the  bosonic atom of mass
$m$  in state $j=1,2$, $V^{(j)}_{\mbox{tr}}(\vec r)$ is external trapping
potential, $\hat V_1$ is the usual nonlinear potential between the atoms
interacting via a contact interaction of strength $g_{lj}= 4\pi \hbar^2
a_{lj}/m$ between two
atoms in states $l$ and $j$, $a_{lj} $ is the scattering length of
atoms in states  $l$ and $j$,  
and $\hat V_2$ is the potential which allows for
the laser-induced transition of an atom from state $1$ to $2$ and vice
versa via a contact interaction of strength $\chi$.

Coupled mean-field GP equations are obtained by replacing the field
operators $\hat \psi_j(\vec r)$ by the wave functions $\psi_j(\vec r)$,
which are the expectation values of the field operators,  in
the Heisenberg equation \cite{l1}:
\begin{equation}
\mbox{i}\hbar \frac{\partial}{\partial \tau} \hat \psi_j(\vec r,\tau) =
\left[
\hat \psi_j(\vec r ,\tau)
,\hat H\right],
\end{equation}
where the time ($\tau$) dependence is explicitly shown and the result is
 \cite{w1,x1,y1}
\begin{eqnarray}\label{cc} \biggr[
&-&\frac{\hbar^2}{2m}\frac{1}{r}\frac{\partial^2 }{\partial
r^2}r + \frac{1}{2}c_jm\omega^2 r^2 +\sum_{l=1}^2
g_{jl}n_l|\psi_l({
r},\tau)|^2 \nonumber 
\\
&-& \mbox{i}\hbar\frac{\partial}{\partial \tau}\biggr]
\psi_j(r,\tau)+\sum_{l=1}^2\chi (1-\delta_{jl}) \psi_l(r,\tau)=0,
\end{eqnarray}
where $\delta_{jl}$ is the Kronecker delta and 
$n_l$ is the number of condensed atoms in state 
$l$.
Here
 we have specialized to the radially symmetric case and have used the
harmonic oscillator trapping potential $V_{\mbox{tr}}^{(j)}(\vec r)=
c_jm\omega^2 r^2/2$ where $\omega$ is the trap frequency and $c_j$ is a
parameter introduced to vary independently the trap potential in each
quantum state.
The normalization
condition for $\chi=0$ is given by 
\begin{equation}
4\pi \int_0^\infty  r^2 dr |\psi_j (r,\tau)|^2= 1. 
\end{equation}

As in Refs. \cite{k1} it is convenient to use dimensionless variables
defined by $x = \sqrt 2 r/a_{\mbox{ho}}$ , and $t=\tau \omega, $
where
$a_{\mbox{ho}}\equiv \sqrt {\hbar/(m\omega)}$, and $
\phi_j(x,t) = x\psi_j(r,\tau ) (\sqrt 2\pi a_{\mbox{ho}}^3)^{1/2}$ . In
terms of these
 variables Eq. (\ref{cc})  becomes \cite{k1}
\begin{eqnarray}\label{e}
\biggr[ &-&\frac{\partial^2 }{\partial
x^2} + \frac{ c_jx^2}{4} +\sum_{l=1}^2 n_{jl}
\frac{|\phi_l({x},t)|^2}{x^2}
-\mbox{i}\xi_j\frac{|\phi_j({x},t)|^4}{x^4}\nonumber \\
&+&\mbox{i}\gamma_j
-  \mbox{i}\frac{\partial
}{\partial t} \biggr]\phi_j({ x},t)+\sum_{l=1}^2\eta (1-\delta_{jl})
\phi_l({ x},t)=0, \end{eqnarray}
where  $\eta= \chi/(\hbar \omega)$ and  
$n_{jl}\equiv 2\sqrt 2 n_l a_{jl}/a_{\mbox{ho}}$ 
 could be negative (positive) when the
corresponding interaction 
is attractive (repulsive). 
In Eqs. (\ref{e}) we  have introduced
a diagonal absorptive i$\gamma_j$ and a quartic three-body term $
-\mbox{i}\xi_j{|\phi_j({x},t)|^4}/{x^4}$
appropriate to study
collapse \cite{f1,g1}. A nonzero $\gamma_j$ allows the possibility of the
absorption of atoms from an external source into the condensate, whereas a
nonzero $\xi_j$ allows for the possibility of ejection of atoms from the
condensate due to three-body recombination. The total number
of atoms in the condensate is not conserved for nonzero
$\gamma_j$ and $\xi_j$.

The fluctuation of the number of atoms in the two states of the condensate 
is best studied
via the quantities
\begin{equation}
N_j \equiv \int_0^\infty |\phi_j(x,t)|^2 dx, \quad j=1,2.
\end{equation}
In the presence of a general absorptive interaction $\eta \ne 0, \gamma_j
\ne 0,$ and $\xi_j\ne 0,$ the reduced number for the two components of the
condensate are given by $n_{11}N_1$ and $n_{22}N_2$, the quantities $N_1$
and $N_2$ carry the information about time evolution of the number of the
two components. The actual number of atoms in the two components are
$n_jN_j$, $j=1,2$. 

For $\gamma_j=\xi_j=\eta =0$, the number of atoms in the two components
are conserved separately. This condition leads to $N_j=1$.
For $\gamma_j=\xi_j=0$ and $\eta\ne 0$, the numbers of atoms in the two
components are not conserved separately $N_j\ne 1$;  
but the  total number is conserved. In 
the presence of absorptive interaction  $\gamma_j\ne 0$ and/or
$\xi_j\ne 0$, there is no conservation of even the  total number
of atoms.

\section{Numerical results}

\subsection{Computational Procedure}

To solve Eqs. (\ref{e})  numerically the proper boundary
conditions as $x\to 0$ and $\infty$ are needed. For a confined condensate
the asymptotic form of the physically acceptable solution is given by 
$\lim_{x\to \infty}|\phi_j(x,t)|\sim \exp (-x^2/4) $.
At $x=0$ one has the regularity condition 
$\phi_j(0,t) = 0$.

Next we discretize Eqs. (\ref{e})  in both space and time by
using a space step $h$ and time step $\Delta $ with a finite difference
scheme using the unknown ${(\phi_j)}^k_p$ which are approximation to
the exact solution $\phi_j(x_p,t_k)$ where $x_p= p h$ and $t_k=k\Delta $.
After discretization we obtain a set of algebraic equations which could
then be solved by using the known asymptotic boundary conditions 
\cite{f1,g1,k1}.  The time
derivative in these equations involves the wave functions at times $t_k$
and $t_{k+1}$. To discretize Eqs.  (\ref{e})
we express the wave functions and
their derivatives   in terms of the average
over times $t_k$ and $t_{k+1}$ \cite{koo} and the result is
the following
Crank-Nicholson-type scheme \cite{koo}  
 \begin{eqnarray} &\mbox{i}&\frac{(\phi_j)_p^{k+1}-(\phi_j)_p^{k} }{\Delta
} =
-\frac{1}{2h ^2}\biggr[(\phi_j)^{k+1}_{p+1}-2 (\phi_j)^{k+1}_{p}\nonumber
\\&+&(\phi_j)^{k+1}_{p-1} + (\phi_j)^{k}_{p+1}-2
(\phi_j)^{k}_{p}+(\phi_j)^{k}_{p-1}\biggr]\nonumber \\
&+&\frac{1}{2}\left[\frac{c_jx_p^2}{4}+\sum_{l=1}^2
n_{jl}\frac{|(\phi_l)_p^{k}|^2}{x_p^2}-i\xi_j
\frac{|(\phi_j)_p^{k}|^4}{x_p^4}+i\gamma_j \right]\nonumber \\ &\times&
\left[(\phi_j)_p^{k+1}+(\phi_j)_p^k\right]+\sum_{l=1}^2\eta(1-\delta_{lj})
{(\phi_l)_p^k}
, \label{fx} \end{eqnarray}
where $j=1,2$.

Considering that
the wave function components $\phi_j$  are  known at time $t_k$, Eqs.
(\ref{fx})  involve 
the unknowns $-$ $(\phi_j)_{p+1}^{k+1},(\phi_j)_p^{k+1}$ and
$(\phi_j)_{p-1}^{k+1}$ at time $t_{k+1}$.
In a lattice of $N$ points Eqs. (\ref{fx})  represent a
tridiagonal set for
$p=2,3,...,(N-1)$ for a specific component $\phi_j$. This set has a unique
solution if the wave functions at
the two end points $(\phi_j)_{1}^{k+1}$ and $(\phi_j)_{N}^{k+1}$ are
known \cite{koo}. These values at the end points are provided by the known
asymptotic boundary conditions.

To solve Eqs. (\ref{fx})  we employ space step $h=$ 0.0001
with $x_{\mbox{max}}\le 15$ and time step $\Delta$ =0.05.
After some experimentation we find that these values
of the steps give good convergence. The iteration is started with the
known normalized (harmonic oscillator) solution of Eqs. (\ref{e}) 
obtained with $n_{jl}=\gamma_j=\xi_j=\eta=0$.  The nonlinear
parameters $n_{jl}$  are increased by equal amounts over
1000 time iterations starting from zero at $t=-50$ until the desired final
value is
reached at $t=0$.  
During the iteration we keep $\xi_j=\gamma_j=0$ and attribute a
small numerical value to $\eta$ ($<0.2$). The resulting
solution is the ground state of the condensate corresponding to the
specific nonlinearity and built up with this particular value of $\eta$.
Then to study the dynamics we allow the system to evolve by continuing the
iteration
with $\Delta=0.05$ and $h=0.0001$ for positive $t$, but maintaining $\eta$ 
and $n_{ij}$ at the
constant final values attained at $t=0$.  
 To study the dynamics for positive $t$ for attractive atomic 
interaction, the absorptive
contact ($\gamma_j$) and  the three-body interactions ($\xi_j$)  
are switched on at $t=0$. This will clearly show the possible collapse 
of the system.

\subsection{Stable Condensate: Repulsive Interaction}

First we  consider the stationary solution of Eqs. (\ref{e}) with
$\gamma_j=\xi_j=0$  in the repulsive case.  We  now choose the
following two sets of 
of  parameters before the  actual calculation:   
$ n_{11}=n_{22}=10,$
$n_{12}=n_{21}=5, $
$c_1=1$, $c_2=0.25, (a) \eta=0.075$ and (b) $ \eta=0.1$.  
This will show the effect of the feedback
between the two atomic states. In this case all interactions are repulsive
corresponding to positive sign of all $n_{jl}\equiv 2\sqrt 2 n_j
a_{jl}/a_{\mbox{ho}}$. 

Although the parameters above are in dimensionless 
units, it is easy to associate them with a physical problem of
experimental
interest in the case $\eta=0$. 
For example, for the mixture of $|F=1,m=-1\rangle $
and $|F=2,m=1\rangle $ states of $^{87}$Rb, the ratio of scattering
lengths $a_{|1,-1\rangle}/ a_{|2,1\rangle} = 1.062$ \cite{u1}. If we label
state $|1,-1\rangle$ by 1 and $|2,1\rangle$ by 2, and consider
$a_{11}/a_{\mbox{ho}} \simeq a_{22}/a_{\mbox{ho}} \simeq
0.002$, then $n_{11}=n_{22}=10$ corresponds to $n_1\simeq n_2\simeq
1770$ for $\eta=0$.  
Hence these models can simulate the actual
experimental
situation composed of two states of $^{87}$Rb.

In Fig. 1  we plot the wave
functions $|\phi_j(x,0)|/x$ for the two components  calculated
with 
two nonzero values of $\eta$ above. In Figs. 2 (a)
and (b) we plot the quantities $N_1$ and  $N_2$ for these models.  When
$\eta=0$, $N_1=N_2=1$. As a nonzero value of $\eta$ is taken, the
quantities $N_j$ oscillate with time
as can be seen from Figs. 2 (a) and (b). When $N_1$ increases, $N_2$ decreases
and this denotes a net transformation of atoms  from  state 2 
to 1 and vice versa. 
These transformations lead to interesting oscillation in $N_j$ and
consequently,  in the actual number of atoms in the two states. 
 For a small
$\eta$ these oscillations are small and as $\eta$ increases these
oscillations increase in amplitude. These oscillations clearly show the
continued dynamical conversion of one species of atoms to another 
 in the otherwise stable condensate.

\subsection{Collapse: Attractive Interaction}

Now we study the simplest case of collapse by taking only the interaction
between the atoms in state 1 to be attractive corresponding to a negative
$a_{11}$. All other scattering lengths $-$ $a_{22}$ and $a_{12}$ (=
$a_{21}$) $-$ are taken to be positive.  Quite expectedly, here the first
component of the condensate could experience collapse. We also consider 
the collapse in both components when all scattering lengths are negative.

The calculation is performed in these cases with  the following
two sets of parameters
(a) $ 
n_{11}=-3.0346, n_{22}=4, $ $n_{12}=n_{21}=1, $
$c_1=0.25$, $c_2=4$,  $\eta=0.2$ and (b)
$n_{11}=-0.67,  n_{22}=-1.4, $
$n_{12}= -0.42, n_{21}=-0.41, $ $c_1=4$, $c_2=0.25$, $\eta=0.1$. 
The wave function components
 at $t=0$ in these cases are
plotted in Fig. 3.  

The above  parameters are in dimensionless units and one
can associate them with a problem of experimental
interest for $\eta=0$
 in terms of two states of $^7$Li. In case of model (a) above 
we consider the state 1 to be the ground
state of
$^7$Li with
attractive interaction as in the actual collapse experiment with
$|a_{11}|/a_{\mbox{ho}}\simeq 0.0005$ \cite{d1,e1}. As $n_{11}=2\sqrt 2
n_1
|a_{11}|/a_{\mbox{ho}}$, this corresponds to a boson number $n_1 \simeq 
2145. $ This number is larger than the maximum number of atoms
permitted in the  BEC of  a single component $^7$Li which is about 1400
\cite{d1,e1}. The state 2 could be one of the excited states of $^7$Li
with
repulsive atomic interaction. The presence of the
second component with repulsive interaction allows for the formation of a 
stable BEC with more
$^7$Li atoms in quantum state 1 than allowed in the single-component BEC.  
We find from Fig. 3 that  $\phi_1$ (full line) is very
much centrally peaked compared to $\phi_2$ (dashed line) in this case. 
This corresponds
to a small
 radius and large central density for the first component  denoting an
approximation to  
collapse. If the number  $n_{1}$ is slightly increased beyond 2200 the
first
component of the condensate wave function becomes singular at the origin
and could experience collapse.

In case of model (b)
we take the atomic
interaction in both
states of $^7$Li  to be attractive corresponding to  negative scattering
lengths. For
$|a_{ii}|/a_{\mbox{ho}}\simeq 0.0005$ as in the actual experiment
\cite{d1,e1}, one has 
$n_1 = 470$ and $n_2 \simeq 990$ for $\eta=0$.  The total number of
particles in this
case is
roughly 1460, which is close  to the critical number 1400 found in
the actual
experiment of collapse in $^7$Li.
Both wave-function
components could become singular in this case as all possible interactions
are attractive. 
The small extension of the wave-function components in space in  this
case as in Fig. 3 denote an approximation to collapse.

Although the collapse of the coupled condensates could be inferred from
the shape of the stationary wave functions of Fig. 3 (sharply peaked
centrally with small  radii), we  also study the dynamics of
collapse from a time evolution of the full GP
equation (\ref{e}) in the presence of an absorption and three-body
recombination, e.g., for $\gamma_j \ne 0$ and $\xi_j \ne 0$ as in the
uncoupled case \cite{f1}. 
The general
nature of time evolution is independent of the actual values of $\gamma_j$
and $\xi_j$ employed provided that a small value for $\xi_j$ and a
relatively larger one for $\gamma_j $  are
chosen \cite{f1}.

First we consider model (a). 
For a dynamic description of this  problem we take for $t>0$
$\gamma_1=\gamma_2=0.04$,
$\xi_1=0.01$ and $\xi_2=0.002$ and allow the solution of the GP equation
to evolve in time using Eq. (\ref{fx}).  In
the presence of these absorptive interactions the total number of atoms  
is not a
constant of motion. The time evolution of $N_1$ and $N_2$  is
shown in  Fig. 4 (a). There is small oscillation in $N_j$ for
$t<0$ signifying a transfer of one type of atoms to the other. For $t>0$
these oscillations are almost undetectable. 
Because of the attractive interaction, the component
1 could experience a succession of collapse. This is manifested in the
successive growth and decay of $N_1$. The condensate corresponding to the
second
component controlled by the
repulsive interaction keeps on growing, manifested by the
continuous growth of $N_2$ with time.

Next we consider the time evolution of the GP equation 
for  model (b) with the following absorptive parameters: 
$\gamma_1=0.04, \gamma_2=0.03, \xi_1=\xi_2=0.005$.
In this case as all interactions are attractive,  both components can
experience collapse. In Fig 4 (b) we plot the evolution of $N_1$ and
$N_2$.
The oscillation of $N_1$ and $N_2$ is clearly visible for
$t<0$. For $t>0$, the absorptive interactions take control and the total
number of atoms start to increase. Consequently, the dynamics of collapse
is more visible than the oscillation in $N_1$ and $N_2$.
Both $N_1$ and $N_2$ is found to exhibit a succession of growth and decay
corresponding to   collapse.
Hence we find that one can have a dynamical collapse of one or both
components of the BEC in the case of a coupled condensate.

\section{Conclusion}

To conclude, we studied the collapse in a trapped BEC of atoms in states 1
and 2 using the GP equation when some of the atomic interactions are
attractive. We include an interaction in the Hamiltonian which allow for a
spontaneous transformation of atoms from state 1 to 2 and vice versa.
Experimentally, this can be achieved by a laser and makes the present
study richer.  In case of all repulsive atomic interactions one can have a
stable condensate of both components. 
When some of  the atomic
interactions are attractive one or both components of the condensate could
experience collapse. The collapse could be predicted
from a stationary solution of the GP equation. The time evolution of
collapse is studied via the time-dependent GP equation with absorption and
three-body recombination.  The number of particles of the component(s) of
BEC experiencing collapse alternately grows and decays with time.  With
the possibility of a detailed study  of a coupled BEC, the results of this
study
could be verified experimentally in the future.

The work is supported in part by the Conselho Nacional de Desenvolvimento
Cient\'\i fico e Tecnol\'ogico  and Funda\c c\~ao de Amparo \`a Pesquisa
do Estado de S\~ao Paulo of Brazil.

\newpage

{\bf Figure Caption:}

1. Wave function components $|\phi_1(x,0)|/x$ (full line) and
$|\phi_2(x,0)|/x$
(dashed line) vs. $x$ for two coupled GP equations with 
$ n_{11}=n_{22}=10,$
$n_{12}=n_{21}=5, $
$c_1=1$, $c_2=0.25, (a) \eta=0.075$ and (b) $ \eta=0.1$.

2. $N_1$ (full line) and $N_2$ (dashed line) vs. $t$
 for models (a) and (b) of 
 Fig. 1.

3. Wave function components $|\phi_1(x,0)|/x$ (full line) and
$|\phi_2(x,0)|/x$
(dashed line) vs. $x$ for two coupled GP equations with 
(a) $ 
n_{11}=-3.0346, n_{22}=4, $ $n_{12}=n_{21}=1, $
$c_1=0.25$, $c_2=4$,  $\eta=0.2$ and (b)
$n_{11}=-0.67,  n_{22}=-1.4, $
$n_{12}= -0.42, n_{21}=-0.41, $ $c_1=4$, $c_2=0.25$, $\eta=0.1$.

4. $N_1$ (full line) and $N_2$ (dashed line) vs. $t$
 for models (a)  and (b) of Fig. 3.


\begin{references}

 \bibitem{a1}J. R.
Ensher, D. S. Jin, M. R. Matthews, C. E. Wieman, E. A.  Cornell, Phys.
Rev. Lett.  {77} (1996) 4984.  
\bibitem{b1}K. B. Davis, M. O. Mewes, M. R.
Andrews, N. J. van Druten, D. S. Durfee, D. M. Kurn, W. Ketterle,
Phys. Rev. Lett.  { 75} (1995)  3969.
\bibitem{c1}  
D. G. Fried, T. C. Killian, L.
Willmann, D. Landhuis, S. C.  Moss, D. Kleppner, T. J. Greytak, Phys. Rev.
Lett.
 { 81} (1998)  3811.


\bibitem{d1}C. C. Bradley, C. A. Sackett, J.
J. Tolett, R. G. Hulet, Phys. Rev. Lett.  {75} (1995) 1687. 
\bibitem{e1}C. Sackett, H. T. C. Stoof, R. G. Hulet, Phys. Rev. Lett. 
{ 80} (1998) 2031.


\bibitem{f1}Yu. Kagan, A. E. Muryshev,  G. V. 
Shlyapnikov, Phys. Rev. Lett.
{ 81} (1998) 933.
\bibitem{g1}
A. Gammal, T. Frederico, L. Tomio,  Ph. Chomaz, Phys.
Rev. A { 61} (2000)  051602. 



\bibitem{h1}
S. Giorgini, L. P. Pitaevskii, S.  Stringari, Phys. Rev. A
{ 54} (1996) R4633.
\bibitem{i1} M. Edwards, P. A. Ruprecht, K. Burnett, R. J.
Dodd,  C. W. Clark, Phys. Rev. Lett. { 77} (1996) 1671. 
\bibitem{j1}S. K. Adhikari,  A. Gammal,
Physica A { 286} (2000) 299.

\bibitem{k1} 
S. K. Adhikari, 
  Phys. Rev.
E
{ 62} (2000) 2937. 




\bibitem{l1} F. Dalfovo, S. Giorgini, L. P. Pitaevskii,  S. Stringari,
Rev. Mod.  Phys. { 71} (1999) 463. 



 \bibitem{m1}R. J. Dodd, M.  Edwards, C. J. Williams, C. W. Clark, M. J.
Holland, P. A. Ruprecht, K. Burnett, Phys. Rev. A { 54} (1996) 661.

 \bibitem{n1} M. Houbiers, H. T.  C. Stoof, {Phys. Rev. A 54} (1996) 5055.

 \bibitem{o1}
 A.  Eleftheriou, K. Huang, {Phys. Rev. A  61} (2000) 043601.
\bibitem{p1}  S.
K. 
Adhikari, Physica A { 284} (2000) 97. 
\bibitem{q1}L. Berg\'e, T. J. Alexander, Y. S. Kivshar, Phys. Rev. A 62
(2000) 023607.


\bibitem{r1}E. P. Gross, Nuovo Cimento { 20} (1961) 454.
\bibitem{s1}
L. P. Pitaevskii, Zh. Eksp. Teor. Fiz. { 40} (1961) 646
[Sov. Phys. JETP { 13} (1961) 451].

\bibitem{t1}
M. R. Matthews, B. P. Anderson, P. C. Haljan, D. S. Hall,
C. E. Wieman,  E. A. Cornell,  Phys. Rev. Lett. { 83} (1999)
2498.
\bibitem{u1} 
M. R. Matthews, D. S. Hall, D. S. Jin, J. R. Ensher, C. E.
Wieman, E. A. Cornell, F. Dalfovo, C. Minniti, S. Stringari, Phys. Rev.
Lett.  { 81} (1998) 243.

\bibitem{v1} C. J. Myatt, E. A. Burt, R. W. Ghrist, E. A. Cornell, 
C. E. Wieman, Phys. Rev. Lett. { 78} (1997)  586.

\bibitem{w1}A. Sinatra, P. O. Fedichev, Y. Castin, J. Dalibard,  G.
V.
Shlyapnikov, Phys. Rev. Lett. { 82} (1999) 251.
\bibitem{x1}
B. D. Esry, C. H. Greene, J. P. Burke, Jr.,  J. L. Bohn,
Phys. Rev. Lett. 
{ 78} (1997) 3594.
\bibitem{y1} T.-L. Ho, V. B. Shenoy, Phys. Rev. Lett. 
 { 77} (1996) 3276.








\bibitem{koo} S. E. Koonin, Computational Physics, Benjamin/Cummings Pub.
Co. Inc., Menlo Park, 1986, pp  161-167.





\end{references}
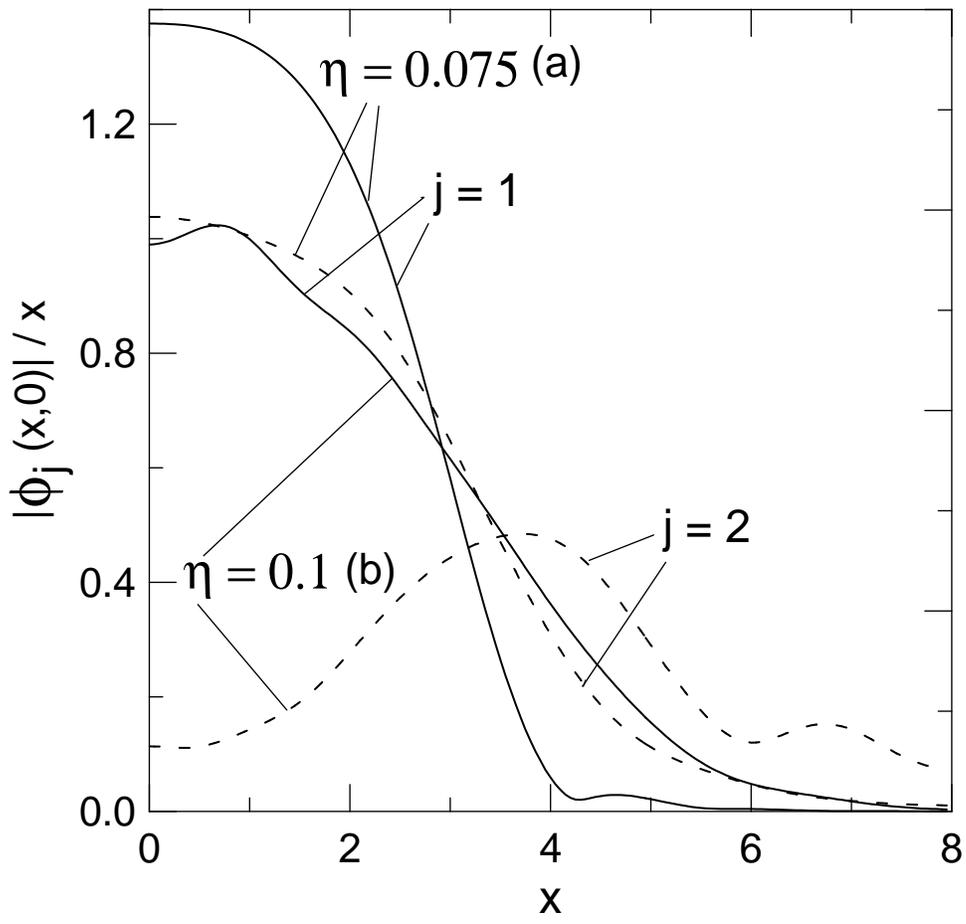
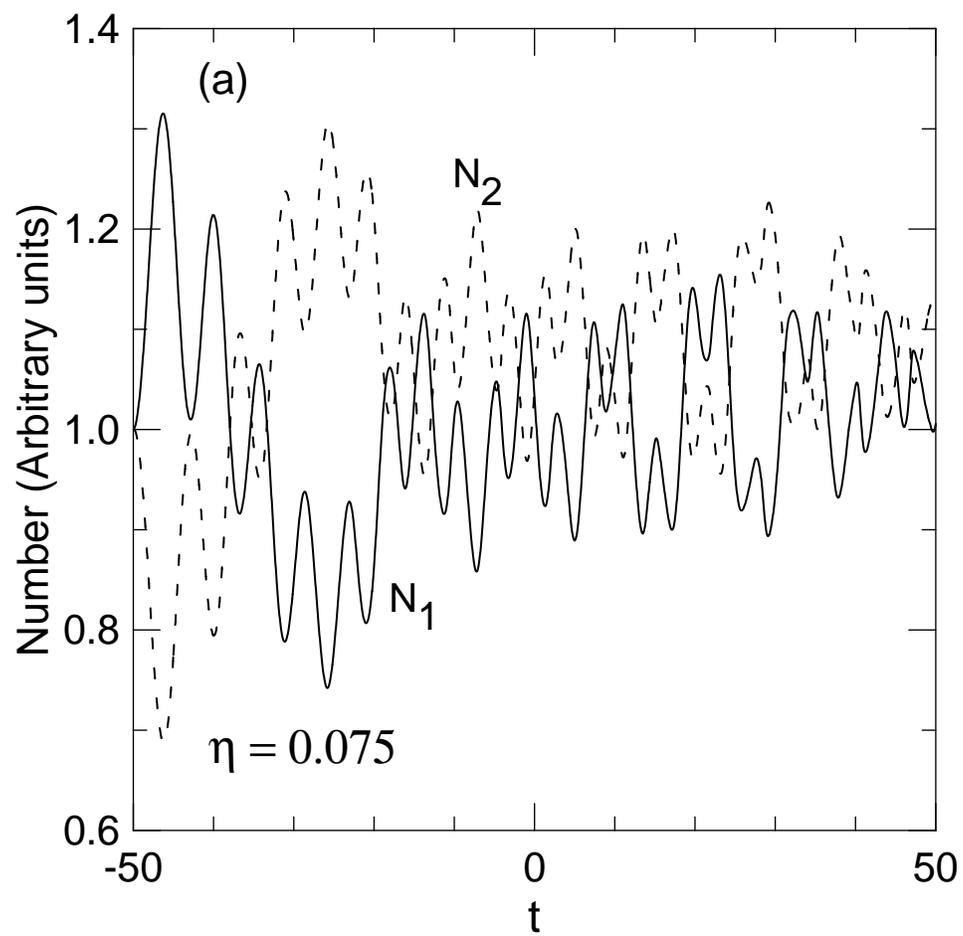
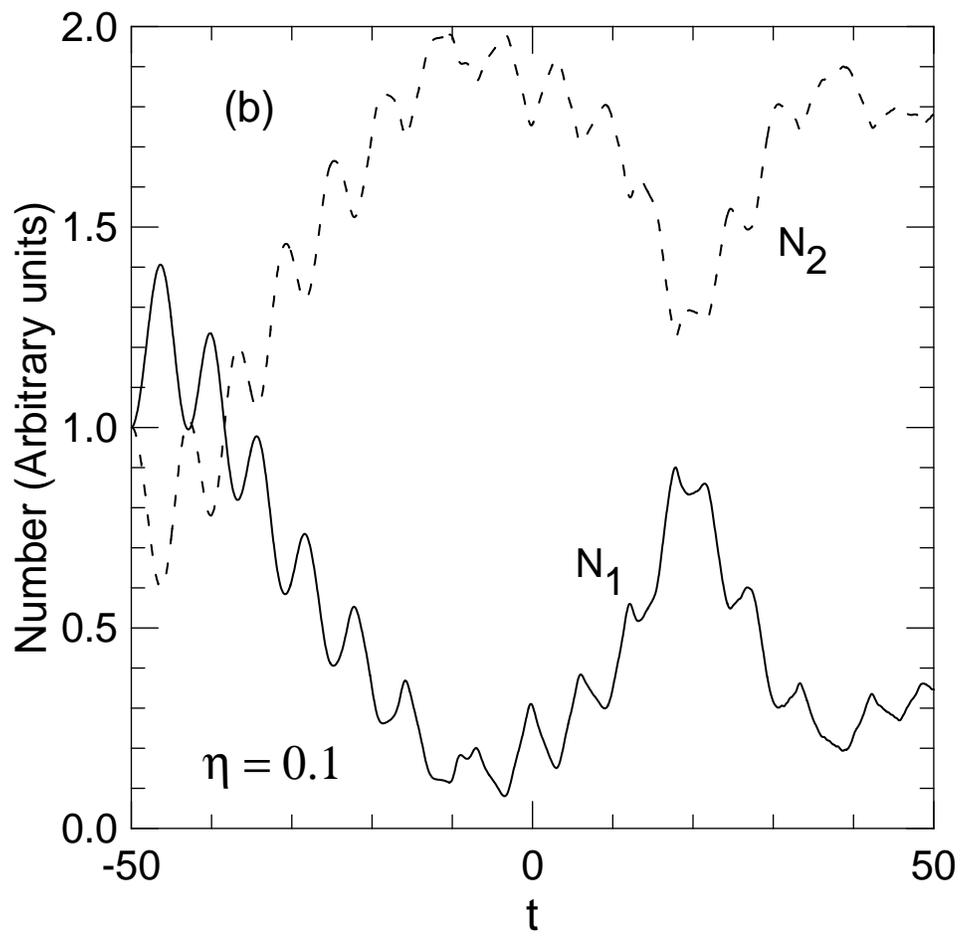
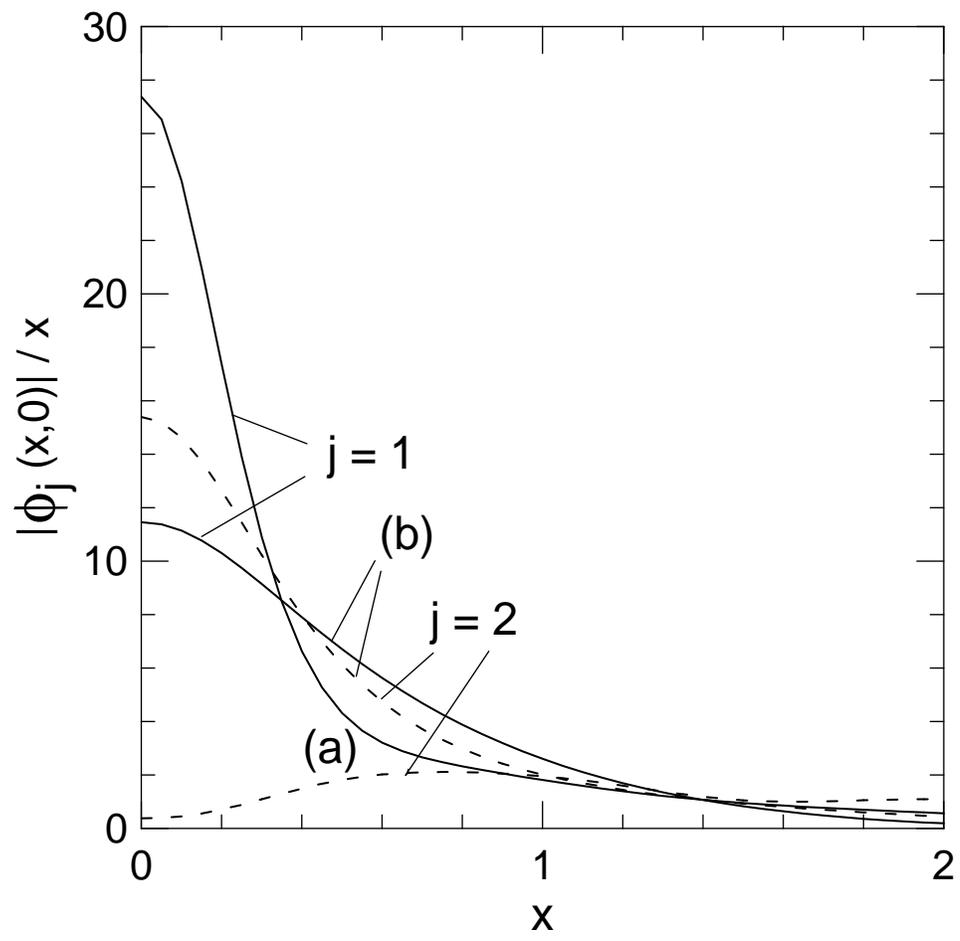
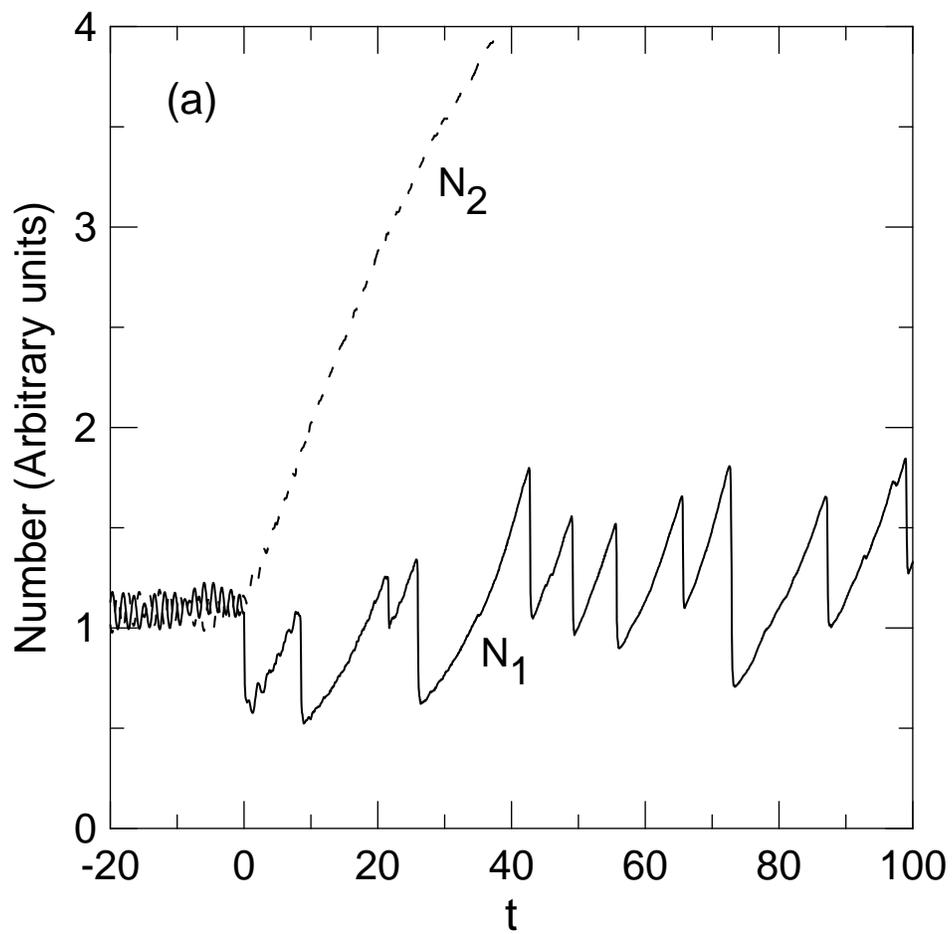
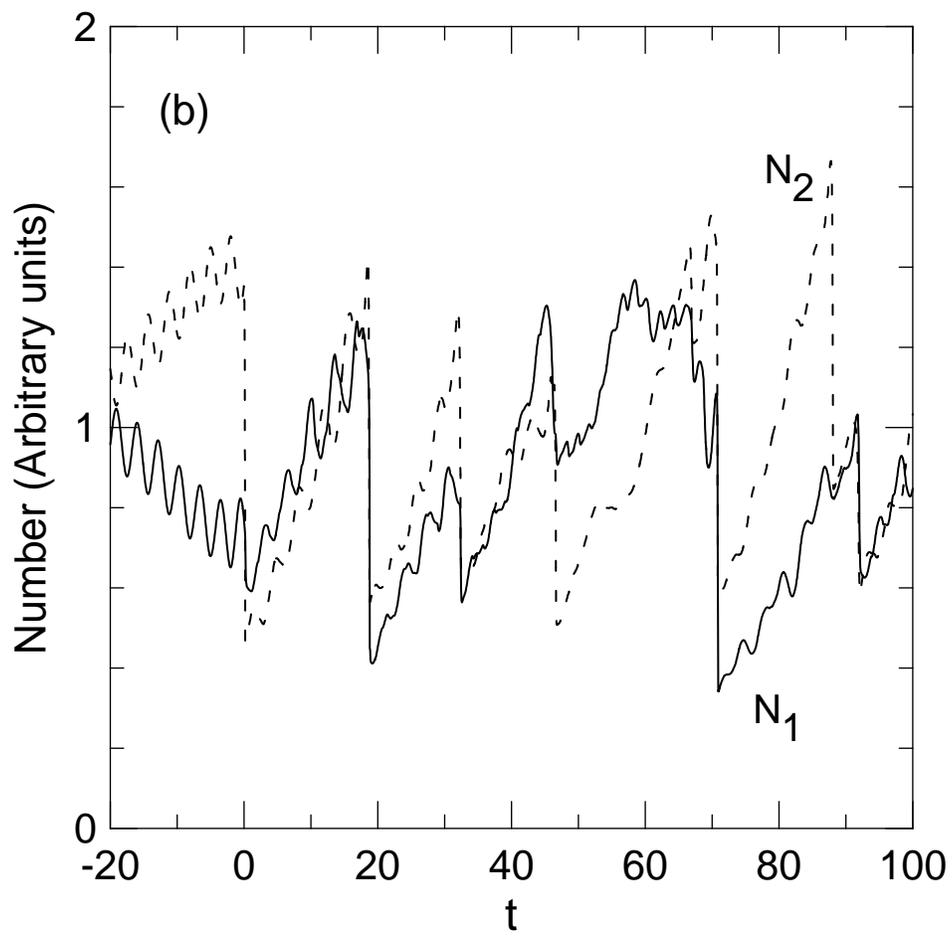
\end{document}